\documentclass[graybox]{svmult}

\usepackage{mathptmx}       
\usepackage{helvet}         
\usepackage{courier}        
\usepackage{type1cm}        
\usepackage{makeidx}         
\usepackage{graphicx}        
\usepackage{multicol}        
\usepackage[bottom]{footmisc}

\makeindex             

\begin{document}

\title*{How to Correctly Stitch Together {\it Kepler} Data of a Blazhko Star}
\titlerunning{How to stitch the data of a Blazhko Star?} 
\author{L. \c{C}el\.{i}k, F. Ekmek\c{c}\.{i}, J. Nemec, K. Kolenberg, J. M. Benk\H{o}, R. Szab\'{o}, D. W. Kurtz, 
K. Kinemuchi, H. V. \c{S}enavc{\i}} 
\authorrunning{\c{C}elik et al.} 

\institute{L. \c{C}elik, F. Ekmek\c{c}i, H. V. \c{S}enavc{\i} \at Ankara Univ., Faculty of Science, Dept. of Astronomy and Space Sciences, 06100, Tando\u{g}an, Ankara, Turkey, \email{lalecelik81@gmail.com}, \and J. Nemec \at Dept. of Physics \& Astronomy, Camosun College, Victoria, British Columbia, V8P 5J2, Canada, \and K. Kolenberg \at Harvard-Smithsonian Center for Astrophysics, 60 Garden St., Cambridge MA 02138 USA,\\
Instituut voor Sterrenkunde, Celestijnenlaan 200D, 3001 Heverlee, Belgium \and J. Benk\H{o}, R. Szab\'{o} \at Konkoly Obs. of the Hungarian Academy of Sciences, Konkoly Thege Mikl\'{o}s \'{u}t 15-17, H-1121 Budapest, Hungary, \and D. KURTZ \at D. Kurtz \at 
Jeremiah Horrocks Institute, Univ. of Central Lancashire, Preston PR1 2HE, \and K. Kinemuchi \at Bay Area Environmental Research Inst./NASA Ames Research Center, MS 244-30, Moffet Field, CA 94035, USA}
%
\maketitle

\abstract*{One of the most challenging difficulties that precedes the frequency analysis of {\it Kepler} data for a Blazhko star is stitching together the data from different seasons (quarters). We discuss the preliminary steps in the stitching, detrending and rescaling process using the data for long-term Blazhko stars. We present the process on {\it Kepler} data of a Blazhko star with a variable Blazhko cycle and some first results of our analysis.}

\abstract{One of the most challenging difficulties that precedes the frequency analysis of {\it Kepler} data for a Blazhko star is stitching together the data from different seasons (quarters). We discuss the preliminary steps in the stitching, detrending and rescaling process using the data for long-term Blazhko stars. We present the process on {\it Kepler} data of a Blazhko star with a variable Blazhko cycle and some first results of our analysis.}

\section{Stitching, Detrending and The Rescaling Process for {\it Kepler}}
\label{sec:1}

Several models have been proposed to explain the Blazhko effect (see e.g., 
\cite{Kolenberg2010}) but it still remains a problem to be solved.  An additional 
difficulty for the analysis of Blazhko stars is that the data obtained in 
subsequent quarters display some discrepancies in their flux values. Therefore, 
when stitching together the light curves from different quarters, these 
discrepancies must be removed.   
 
\begin{figure} 
\includegraphics[scale=0.55]{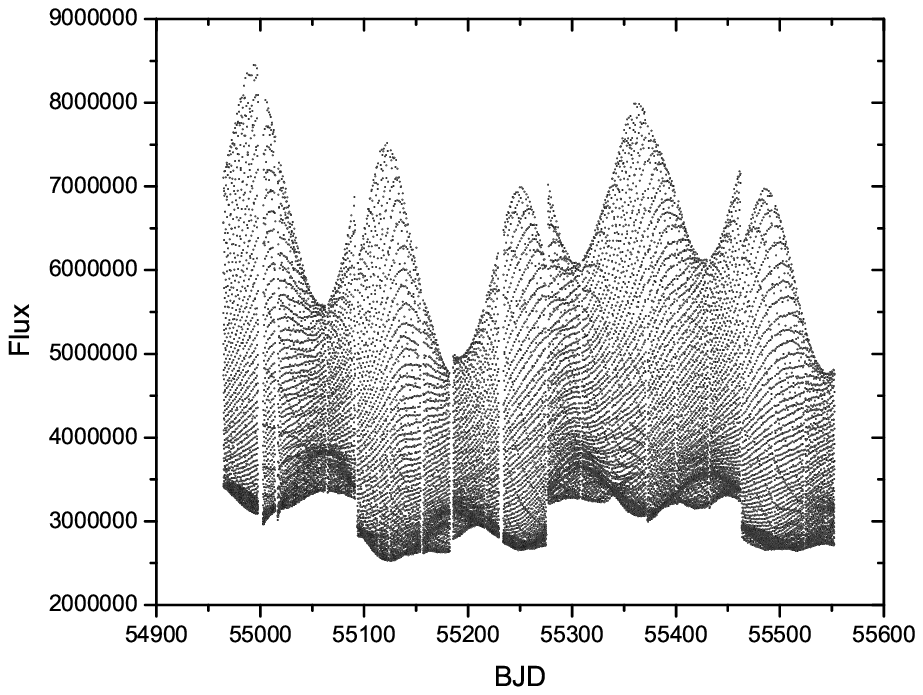}
\includegraphics[scale=0.55]{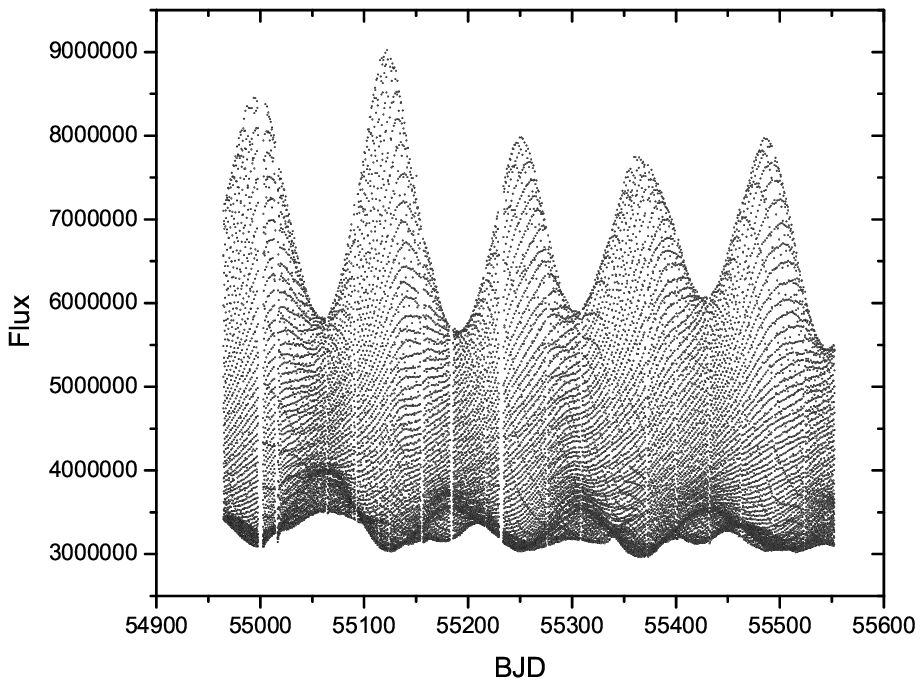}
\caption{The light curve of a Blazhko star, using the data from Q1 up to Q7 
quarters without rescaling procedure (left panel), and with rescaling procedure 
(right panel)} 
\label{fig:1} 
\end{figure}

To overcome the problems originating from {\it Kepler} itself and/or from the 
``Automated Pipeline" routine, the users of the {\it Kepler} archive can use the 
PyKE\cite{PyKE} software. In this study, we applied the rescaling process to five 
Blazhko stars. After stitching the data for all quarters, the most notable 
difference is the flux offset between subsequent quarters originating mainly from 
the instrumental effects (see left panel of Fig.~\ref{fig:1}). Our rescaling 
process matches the light curves from consecutive quarters. This matching is based 
on the assumption that phase-ordered light curves with a few cycles closest to 
each other between two consecutive quarters must have nearly the same flux values 
at the same phases. The first parameter is the period of the star. Another 
parameter, the folding epoch, must be determined to carry out the phase ordering 
process. During the matching process, the corresponding flux values for the same 
phases between phase-ordered light curves of two consecutive quarters are 
determined and proportioned. Therefore, the phase scaling factors are determined 
for, and applied to short ranges of phase. The right panel of Fig.~\ref{fig:1} 
represents a simple diagram of this approach for the rescaling process.

\begin{acknowledgement}
We thank M. E. T\"{O}R\"{U}N (MSc) for his assistance during the software improvements and thank the entire {\it Kepler} team for the efforts which have made these results possible. 
\end{acknowledgement}

\end{document}